\documentstyle[aasms4,flushrt]{article}

\newcommand{\ii}{\'{\i}}
\newcommand{\beq}{\begin{equation}}
\newcommand{\eeq}{\end{equation}}
\newcommand{\beqa}{\begin{eqnarray}}
\newcommand{\eeqa}{\end{eqnarray}}

\newcommand{\lsim}{\mathrel{\hbox{\rlap{\hbox{\lower3pt\hbox{$\sim$}}}\lower-2pt\hbox{$<$}}}}
\newcommand{\gsim}{\mathrel{\hbox{\rlap{\hbox{\lower3pt\hbox{$\sim$}}}\lower-2pt\hbox{$>$}}}}

\journalid{.......}{.......}
\articleid{.......}{.......}
\paperid{}
\slugcomment{Submitted to Astronomical Journal. }

\begin{document}

\title{A Search  for Old Star Clusters in the Large Magellanic Cloud}

\author{Doug Geisler}
\affil{Kitt Peak National Observatory, National Optical Astronomy Observatories,\\
P.O. Box 26732, Tucson, Arizona 85726}

\author{Eduardo Bica, Horacio Dottori}
\affil{Departamento de Astronomia, Instituto de F\'{\i}sica, UFRGS,  
        C.P. 15051, 91501-970  \\
Porto Alegre RS, Brazil}

\author{Juan J. Clari\'a\altaffilmark{1}, Andr\'es E. Piatti\altaffilmark{1}}
\affil{Observatorio Astron\'omico de C\'ordoba, Laprida 854, 5000, \\ C\'ordoba,
Argentina}

\author{and}

\author{Jo\~ao  F. C. Santos Jr.\altaffilmark{2}}
\affil{Departamento de Astronomia, Instituto de F\'{\i}sica, UFRGS,  
        C.P. 15051, 91501-970  \\
Porto Alegre RS, Brazil}

\altaffiltext{1}{ Visiting Astronomer, Cerro Tololo Inter-American
Observatory, which is operated by AURA, Inc., under cooperative
agreement with the NSF}
\altaffiltext{2}{ Present Address: Dep. de F\ii sica, ICEx, UFMG,
C.P. 702, 30123-970 Belo Horizonte MG, Brazil}

\begin{abstract}

There are only a handful of known star clusters in the LMC that are genuinely old,
i.e. of similar age to the globular star clusters in the Milky Way. We report
the first results of a color-magnitude diagram survey of 25 candidate old LMC clusters,
which were
uncovered by means of integrated UBV photometry and  CaII triplet spectroscopy
during previous investigations.
The photometry was carried out with the Washington system $C,T_1$ filters on the 
Cerro Tololo 0.9m telescope.
For almost all of the sample, it was possible to reach the turnoff region,
and in many clusters we have several magnitudes of the main sequence.
The efficiency  and efficacy
of the technique are demonstrated by our deep CMD for
ESO 121-SC03 (used as a control and calibrator), which clearly
shows a magnitude of main sequence for this $\approx9$ Gyr old object in a total 
of $<1$ hour of integration time.
Age estimates based on the magnitude difference $\delta T_1$ between the giant 
branch clump and the
turnoff, calibrated using standard clusters,
revealed that no new old clusters were found. The candidates 
turned out to be of intermediate age (1-3 Gyr)
(we cannot rule out old ages for NGC\,1928 and NGC\,1939 since the
turnoff was not reached  for these compact clusters in crowded bar fields).
We show that the apparently old ages as inferred from integrated UBV colors 
can be explained by a combination of stochastic effects 
produced by bright stars and by photometric errors for faint clusters lying in 
crowded fields.  
The relatively metal poor ($[Fe/H]\sim -1.0$) candidates from the CaII 
triplet spectroscopy also turned out 
to be of intermediate age. This, combined with the fact that they 
lie far out in the disk, yields
interesting constraints regarding the formation and evolution of the 
LMC disk. We also study the age distribution of intermediate age and old clusters
considering not only the present $\delta T_1$ parameter, but also $\delta V$ and 
$\delta R$ measured in CMDs from the literature. This homogeneous set of accurate
relative ages allows us to make an improved study of
the history of cluster formation/destruction
for ages $>1$\,Gyr. We confirm previous indications that there was apparently no 
cluster formation in the LMC during the period from 3-8 Gyr ago, and that there was
a pronounced epoch of cluster formation beginning 3 Gyrs ago that peaked at about 1.5 
Gyrs ago.
Our results suggest that there are few, if any, genuine old clusters in the LMC 
left to be found. 
\end{abstract}

\section{Introduction}
\label{intro}
Globular cluster systems are tracers of the oldest stellar populations in galaxies,
and the determination of their  number and the study of their properties are fundamental to  understanding
the formation and evolutionary  processes in  galaxies. The  largest globular cluster systems 
occur in giant ellipticals, often with populations exceeding 
$10^3$ clusters (see Harris 1991 for a review).
Battistini et al. (1993) estimated a best sample  of 341  globular clusters and candidates in 
M\,31. Webbink (1985) compiled a list of 154 globulars in our own Galaxy; however, 
subsequent observations  have revealed that several of these are background galaxies, 
planetary nebula in rich fields or open clusters
(e.g. Djorgovski et al. 1990, Bica et al. 1995) while some new members have
been added, including  Lyng\aa\,7 (Ortolani et al. 1993),
Pyxis (Da Costa 1995, Irwin et al. 1995, 
Sarajedini \& Geisler 1996) and IC 1257 (Harris et al. 1997).

As reviewed by Suntzeff et al. (1992),
the Large Magellanic Cloud has eight bonafide
old clusters (here taken to mean similar in age to globular clusters in the Galaxy),
including NGC\,1466, 1786, 1835, 1841, 2210, 2257, Hodge\,11 and Reticulum, confirmed by means
of deep color-magnitude diagrams (CMDs) and/or the presence of RR-Lyrae stars.
Five other clusters (NGC\,1754,
1898, 1916, 2005 and 2019) have long been suspected of being old on the basis of 
crude CMDs,
integrated properties and/or  metallicity estimates (e.g. Hodge 1960, Olszewski et al. 1991)
but require high quality  CMDs for a definitive 
diagnosis, since they are compact and/or embedded in dense fields.
Such CMD investigations using HST are currently underway.

LMC clusters with ages determined from CMDs show a pronounced
gap between a large number of intermediate age 
clusters (age $\sim 1-3$ Gyr, hereafter IACs)
and the classical globular clusters, with ages $>12$ Gyr (Olszewski et al. 1991), 
with the sole exception of 
ESO\,121-SC03 with an age of $\sim 9$ Gyr (Mateo et al. 1986). 
As emphasized by Da Costa (1991), this prevents us from using the known clusters to
tell us any details about the chemical evolution of the LMC over most
of its history, despite our ability to determine accurate ages and metallicities
for them (Olszewski et al. 1991). Finding even a single additional
cluster in this age gap, or more bonafide
old GCs, would significantly enhance our knowledge of the chemical evolution
of this important galaxy.
Other fundamental questions to be addressed by this study 
are: (i) Is the present known sample of  old clusters  complete? 
A comparison of the luminosity function for old globular clusters in the Galaxy and LMC
suggests  that the LMC may be harboring several faint old clusters previously unrecognized
(Suntzeff et al. 1992); and
(ii) Have less massive old clusters formed and dynamically survived 
in the LMC, similar to the Palomar globular clusters in the Galaxy?
In order to assess these issues we have undertaken a systematic CMD survey
of old
globular cluster candidates which have been recently published.

Several studies have uncovered a number of interesting old cluster candidates.
Our primary source of candidates was the catalog of integrated UBV photometry for 
624 star clusters and associations in the LMC by Bica et al. (1996).  They
derived equivalent SWB types (Searle et al. 1980) for these clusters from their 
location in the  (U-B) vs. (B-V) diagram. The SWB sequence is related primarily
to age (e.g. Elson \& Fall 1985, Chiosi et al. 1988).
In particular, the SWB type VII clusters are known to be comparable
to the Galactic GCs in age and abundance. There are 37  clusters in the Bica et al.
sample that fall in the SWB type VII domain, 22 of which had no previous CMD 
available or were 
not scheduled for observation with the HST at the time of our observing run.
However, it is known that such integrated data is not foolproof in isolating
clusters older than 3 Gyr (Chiosi et al. 1988), so that CMDs are required
for a definitive diagnosis (Bica et al. 1996).

A second source of candidates are 4 clusters (3 of them observed in the
present work) from Olszewski et al. (1991)
which have metallicities $-1.2<$[Fe/H]$<-0.8$ but have no further information
regarding their age, including SWB type. The age-metallicity relation derived by Olszewski
et al. for a large number of LMC clusters suggests that any cluster with such 
a relatively low metallicity
is an excellent candidate for type VII membership. For example, the lowest abundance
for an IAC  is $\sim -0.7$, while ESO 121-SC03 has a metallicity
of $-0.9$. Thus, these 4 candidates are about as likely as the above sample 
to provide genuine old clusters.

In Section\,2 we present the observations and reduction procedure. In Section\,3
we determine ages based on the magnitude difference between the 
giant branch clump and the main sequence turnoff. We also discuss the impact of the
new clusters on our knowledge of the age distribution of LMC clusters. 
In Section\,4 we
study stochastic effects of the number of bright stars
on the integrated colors, as well as the influence of photometric errors, in 
particular for less massive clusters. In Section 5 we discuss new insights into the 
chemical enrichment  of the LMC afforded by our results.
Section 6 presents concluding remarks.

\section{Observations and Reductions}

Twenty three candidate old LMC clusters from Bica et al. (1996) and Olszewski et al. (1991)
were observed with the CTIO 0.9m telescope
in December, 1996 with the Tek2k \#3 CCD. 
We also observed ESO121-SC03 as a reference cluster in order to test how
deep this instrumental configuration would reach, and
as an age calibrator. 
The scale on the chip is $0.40\arcsec$ per pixel, and consequently the
area covered by a frame ($2048^2$ pixels) is about 
$13.6\arcmin\times13.6\arcmin$.
Two additional Bica et al. clusters (NGC2153 and SL769) were observed with the
CTIO 4m in February, 1996 with the Tek2k \#4 CCD, with similar pixel and areal coverage.
For both observing runs, we obtained
data in the Washington (Canterna 1976)
C and Kron-Cousins R filters. Geisler (1996) has shown that the
latter is a more efficient substitute for the standard Washington T$_1$ filter, and we 
used his prescription for a new C filter which is both more efficient and also has a
much smaller color term than previous C filters. We elected
to use the Washington system because of its combination of broad
bands, and high metallicity sensitivity provided by the C filter and the
wide color baseline
between C and T$_1$. These data, in addition to providing us with age determinations,
will also allow us to derive accurate metal abundances 
based on the standard giant branch technique outlined in Geisler and Sarajedini (1996;
1997), and will be the subject of another paper in this series.

We obtained generally a single 15 minute T$_1$ (R) exposure and a 45 minute C exposure
per cluster. Thus, only an hour of total integration time with an 0.9m telescope was 
spent on each cluster.
The observed clusters are listed in Table 1, together with their integrated
V magnitudes (Bica et al. 1996) and their metallicities (Olszewski et al. 1991),
when available. In addition to the clusters in the Table, we also observed IC\,2134
(SL\,437, LW\,198) and SL769 which will be discussed in detail elsewhere, due to 
apparently composite stellar populations in the CMD. 
Figure 1 shows the $T_1$ frame of the SL555 field, which is typical of the 
magnitude  of the cluster sample and the crowding of the field, although our sample
includes large ranges in both of these quantities.

The data were reduced with the stand-alone version of the DAOPHOT II program (Stetson 1987)
after standard trimming, bias subtraction and flat-fielding.
A quadratically varying PSF yielded aperture corrections which were essentially constant
with position. The standard FIND-PHOT-ALLSTAR procedure was performed 3 times on each
frame, with typically 30000 objects being measured.
Mean aperture corrections were generally determined to only 0.01 mag (rms) in both 
filters in all but the most crowded frames. 

In order to standardize our photometry, standard stars from the lists of Harris and 
Canterna (1979) and Geisler (1996) were observed at the beginning, middle and
end of each night. A mean of 31 standards was observed per night. Although the air
mass range of the standards was rather small, it did encompass the range in which our
program clusters was observed. Transformation equations of the form:

\noindent $c = C + a_1 + a_2\times (C-T_1) + a_3\times X_c$;~~~~~~~~~~~~~~~~~~~~~~~~~~~~~~~~~~~~~~(1)

\noindent $t_1 = T_1 + b_1 + b_2\times (C-T_1) + b_3\times X_{T_1}$~~~~~~~~~~~~~~~~~~~~~~~~~~~~~~~~~~~~~(2)

\noindent were used, where c and $t_1$ refer to instrumental magnitudes (corrected to 1s
integration using a zero point of 25.0 mag), C, $T_1$ and $(C-T_1)$ are standard values, and
the appropriate air masses are given by X. We first solved for all three transformation
coefficients simultaneously (using the PHOTCAL package in IRAF
\footnote{IRAF is distributed by the National
Optical Astronomy Observatories, which is operated by the Association of Universities
for Research in Astronomy, Inc., under cooperative agreement with the National
Science Foundation}) for each night
and found mean color terms
of $-0.073 \pm 0.003(\sigma)$ in c and $-0.012 \pm 0.002$ in $t_1$ for the 5 nights. 
Note that the c color term in the revised $C$, $T_1$ system (Geisler 1996) is 2-3 times
lower than typical values obtained with previous C filters, as designed.
We then substituted these
mean color terms into the above equations and solved for the remaining two coefficients
for each night simultaneously. Typical values were 2.2 and 1.8 for the c and $t_1$
zero points, and 0.45 and 0.3 for the c and $t_1$ airmass coefficients. The nightly rms
errors from the transformation to the standard system ranged from 0.018 to 0.024 in c and
0.010 to 0.020 in $t_1$, with means of 0.021 and 0.014, indicating the nights were all of
excellent photometric quality.

We then applied these transformation equations to the results of the final ALLSTAR run
after first applying the appropriate aperture correction.
The photometry for the individual clusters will be presented in subsequent papers. It is
currently available from the first author upon request.

Our CMD for ESO121-SC03 is shown in Figure 2. This object was included in
our sample because it is the unique cluster within the age gap between the IACs and
classical old globulars in the LMC, as found by Mateo et al. (1986), and thus serves as
a test of how well our technique works. Clearly, we have reached to about a magnitude
below the main sequence turnoff in this $\approx9$\,Gyr cluster and we can therefore be confident
that our technique should easily be able to distinguish IACs and older clusters for all
but the most crowded fields.
Indeed, our CMD is very comparable to that of Mateo et al. obtained with the CTIO 4m but
with a less sensitive chip. The horizontal branch is very clearly defined as is 
the turnoff.
Note that an equal-area field contains typically only 14 stars, which lie either at
very red colors or near the main sequence. A statistical subtraction of these objects
from the cluster CMD would not significantly change its appearance or the parameters
we derive.

\section{Ages}
\label{sec:age}

The  magnitude difference ($\delta$) between the giant branch clump in intermediate age
clusters (the horizontal branch in old clusters) and the main sequence turnoff
is well correlated with age (e.g. Phelps et al. 1994 and references therein). 
Phelps et al. 1994 and Janes \& Phelps (1994) have studied a large sample of
intermediate age Galactic open clusters with  BV CMDs. 
They concluded that this magnitude difference (as well as the color difference of these 
features) provides very accurate relative ages, especially for IACs,
since it is independent of reddening and  distance modulus. The difference
is  virtually  independent of absolute calibrations, and avoids isochrone
fitting which can be very model dependent for IACs.

 Most of the available CMDs of LMC clusters are $V vs. (B-V)$, with some   
$R vs. (B-R)$ as well. We list in Table 2 all LMC IACs with CMDs
reaching the main sequence turnoff, which are not expected to be seriously affected
by field contamination.
We also include some blue clusters (SWB\,III) and  known
old clusters (SWB\,VII) for comparison purposes. Available integrated V magnitudes and equivalent SWB types
are from Bica et al. (1996), and metallicities from Olszewski et al. (1991).
We measured $\delta$V and $\delta$R for the literature CMDs together
with the turnoff magnitudes, and present them in Table\,2 as well. Typical errors in
these measurements are $\pm$0.15-0.25\,mag. For the present cluster
sample, we measured $\delta{T_1}$ values which are presented in Table\,1. 
Note that this should be within $\sim 0.02$ mags of $\delta R$ given the similarity 
of the two filters (Geisler 1996).
Typical errors are as above. The clusters NGC\,1928 and NGC\,1939 are so compact
and embedded in such dense bar fields that the photometry is not as deep as
for the rest of the sample, and we do not detect the main sequence.
Thus, we cannot infer their ages from the
present data. Two other clusters, SL\,244 and SL\,359, are located in dense
LMC disk fields, so that the CMDs are very contaminated, and the 
$\delta\,T_1$ values (and derived ages) should be taken as lower limits
(Table\,1).

We then tested the similarity of the
$\delta$V\, and \,$\delta$R (or $\delta T_1$) parameters for the clusters which have observations 
in both filters.
One LMC cluster in the literature, NGC\,2213, has
been observed in both $\delta R$ and $\delta T_1$: 
the value  $\delta\,R = 1.3$ (Table\,2) is 
essentially the same, within the uncertainties, as the  $\delta\,T_1 = 1.2$ that
we measured in the CMD of Geisler (1987). In addition to the $\delta$V\,
vs \,$\delta$R values shown in Fig.\,3, we included in the comparison 
two clusters (SL\,842 and ESO\,121-SC03) which have $\delta$\,V (Table\,2)
and $\delta\,T_1$ from the present data (Table\,1). Finally, genuine
old clusters in the LMC have only been observed in BV (Table\,2); in
order to test old clusters in Fig.\,3 we compared the average of the latter
values ($<\delta{V}>$ = 3.4$\pm$0.1) with the average $<\delta\,R>$ = 3.3$\pm$0.2 of the 
Galactic globular cluster Pyxis (from Da Costa 1995, Irwin et al. 1995 and Sarajedini \& 
Geisler 1996). Some additional points in Figure 3 were measured from
CMDs of Galactic open
clusters, for which $\delta V$ and CMD references are: Tombaugh\,2 (1.4, Kubiak 
et al. 1992), M\,67
(2.2, Montgomery et al. 1993), NGC\,1245 (0.4, Carraro \& Patat 1994) and NGC\,6791
(2.9, Montgomery et al. 1994). For the same clusters $\delta T_1$ and references are
given in the discussion of Figure 5 below. A good
correlation is observed in Fig.\,3. In fact, a linear least-squares fit gives 

\noindent $\delta R = \delta T_1 = 0.15 + 0.99\times \delta V$~~~~~~~~~~~~~~~~~~~~~~(3)

with correlation coefficient of $r=0.993$.

We show in Fig.\,4  $\delta\,V$ as a function of the V turnoff magnitude, where a
good relation is seen.
Since the turnoff magnitude is an excellent age indicator, so
are also  $\delta\,V$,  $\delta\,R$
and $\delta T_1$. SWB types are also indicated in the Figure.  
For the range $0.3<\delta\,V<1.2$ the relation is bivalued, and it might be used as an age 
estimator only when combined with additional information, e. g. integrated color.
It is worth noting that the relation presents a minimum for SWB\,IVA.

\subsection{Ages From $\delta T_1$}

For the clusters in our sample,
we first derive an age calibration based on clusters which have $\delta T_1$ values
available, as well as accurate ages. There are 6 such clusters: NGC 2213  
($\delta T_1$=1.2 mag, age = 1.5 Gyr, Geisler 1987) and ESO121-SC03 (3.0 mag - this
paper, 9 Gyr - Mateo et al. 1986) in the LMC,
and the Galactic open clusters NGC 1245 (0.5
mag, 1 Gyr - Wee and Lee 1996), Tombaugh 2 (1.5 mag,
2 Gyr - Wee, Lee and Geisler 1996), M67 (2.4 mag - Geisler and Sarajedini 1997, 4 Gyr -
Dinescu et al. 1995) and NGC 6791 (3.2 mag - Geisler and Sarajedini 1997, 10 Gyr - Tripicco
et al. 1995). Figure 5 plots these points together with  a least-squares fit which was 
obtained by means of the IRAF routine Fitparam. The resulting equation is 

age (Gyr) = 0.23 + 2.31$\times \delta T_1 -1.80\times \delta T_1^2 + 0.645\times
\delta T_1^3$  ~~~~~~~~~~~~~~~~~~~~~~~(4),

\noindent with an uncertainty of $\sim 0.4$ Gyr over the age range of IACs.
We note that a relation of the form: age $\propto 10^{(a\delta T_1 + b\delta T_1^2)}$,
as used by Janes and Phelps,
did not fit the data as well as this relation. 
We then applied equation 4 to our program sample and the results are given in column 5 
of Table 1.

Our intriguing result is that all
the old cluster candidates where we could determine 
$\delta T_1$ turned out to be IACs, with ages between $\sim 1-3$ Gyr.
Figures 6 and 7 show two of our CMDs, including one cluster (H7, 1.4\,Gyr)
in the middle of the age range of our sample
and a cluster amongst the oldest in our sample (SL388, 2.6\,Gyr).
The increasing difference between the clump and the turnoff is obvious, but even in
SL388 this difference is still $\sim $1 mag less than that in ESO121-SC03.
In Section 4 we discuss several likely explanations for why
the candidates from the $UBV$ integrated
photometry turned out not to be old clusters. 
As for the three candidates  from the Olszewski et al. (1991) sample, selected 
because of their relatively low metallicities ([Fe/H]$\approx$-1.0),
they too turned out to be IACs.
These will be discussed further in Section 5.

Two of the integrated $UBV$  candidates 
for old clusters have not been observed in the present 
study. SL\,354 (LW 177) has a preliminary BV CMD (Olszewski 1988)
indicating intermediate age (Table 2). NGC\,1997 (SL\,520, LW\,226) does not have a 
published 
CMD but the high metallicity from the CaII spectroscopy (Olszewski et al. 1991)
also points to an intermediate age.

\subsection{Ages For the Combined Sample}

Using equation (3) it is possible to transform $\delta V$ into equivalent 
$\delta T_1$ values, and by means of the age calibration for $\delta T_1$ (equation (4)), 
to estimate homogeneous relative ages for the literature sample of CMDs in 
Table 2.
For the sample of CMDs involving the $R$ band, $\delta R$ values are used 
directly in equation (4), due to the equivalence with $\delta T_1$. The 
resulting ages are provided in column 11 of Table 2. For comparison purposes
we also give in Table 2 (column 10) the age values derived in the original CMD source
papers. The overall agreement is good. However, since in the CMD sources
different sets of isochrones and different fitting criteria were used, the
magnitude difference  ($\delta$) between clump/HB and turnoff method is
more homogeneous, and consequently the relative ages in the present study 
are expected to be more precise. 

We show in Figure 8 the age histograms for the IACs and old clusters 
(including ESO121-SC03) from both Tables 1 and 2
using the presently derived ages. Notice that the
age scale is linear and the bins are 0.2 and 1\,Gyr for IACs and
old clusters, respectively. This Figure synthesizes all age information
that can be derived from CMDs 
available in the literature for LMC clusters attaining the main sequence
turnoff, in conjunction with the 
present sample. We are possibly witnessing the structure of the
1-3\,Gyr cluster forming epoch, assuming that cluster disruption effects
(Wielen 1988) are negligible. In such a case, the IAC distribution in 
Figure 8 would suggest 
that a burst of star formation occurred, with onset at 
$\approx3$\,Gyr and a peak at $\approx1.6$\,Gyr. We cannot apply the present
age calibration for ages $<1$\,Gyr and consequently incompleteness may affect
the distribution at $\approx1$\,Gyr. 
Alternatively, assuming a constant star formation rate of 10 clusters per
200\,Myr (peak of the distribution), the IAC
distribution between 1.5 and 3\,Gyr would imply a cluster disruption rate
of 77\% per Gyr. 

Our results confirm previous findings (e.g. Da Costa 1991)
that the LMC apparently did not 
form clusters between 3--8 Gyrs ago, a period covering more than $1/3$ of
its lifetime, and that it only formed a single (surviving)
cluster between 3 Gyrs and the epoch of formation of the classical globular clusters.
We have also found additional evidence that the onset of IAC formation
ocurred at 3 Gyr. These results are also in very good agreement with studies of
the star formation history of the LMC field, which find an intense star formation
event ocurred $\sim 2-3$ Gyrs ago, with little star formation before that time
(Butcher 1977, Gallagher et al. 1996, Elson et al. 1997). This argues against the
possibility that the lack of older clusters is due to disruption. 

The sample of old clusters is too small for a discussion of the distribution.
However, it is worth noting that the $\delta$ method detects significant
age differences  between ESO121-SC03 and the older clusters. Also note that we 
have extrapolated our calibration to derive ages for these clusters so their
ages are more uncertain.

\section{Stochastic Effects on Integrated Colors}
\label{sec:stoc}

 The brightest LMC clusters ($10<V<12.5$)
were used by Searle et al. (1980) to define the SWB
zones in a color-color diagram based on the Gunn system. Bica et al. (1992) 
have transposed this (mostly age) sequence to the integrated (U-B) vs (B-V) diagram, 
creating borderlines which encompassed the known SWB clusters. Bica et al. (1996)
determined equivalent SWB types for a large sample, including clusters fainter than
V=14. Since the number of stars in different evolutionary stages
in the CMDs are subject to stochastic effects, fluctuations in integrated
colors are expected, and should become increasingly important for fainter
clusters as the number of bright stars decreases and the relative importance of statistical
fluctuations increases.

  Stochastic effects on optical colors have been previously studied by  King (1966),
Barbaro \& Bertelli (1977), Chiosi et al. (1988), Girardi \& Bica (1993) and Girardi et al. (1995).
In Girardi \& Bica (1993) special attention was devoted to red supergiants, which produce 
large color dispersions for SWB\,I clusters. More recently 
Santos \& Frogel (1997) have studied such effects in  optical and
infrared integrated colors, and found that significant color dispersions in JHK
arise at different ages due to  AGB and red supergiant
stars even for massive clusters.  

  In the present study we  carried out experiments on integrated UBV color fluctuations 
due to Poissonian noise in the number of bright stars in different parts
of the CMD of
individual clusters. We tested stars in the cluster turnoff,  clump and  giant branch 
(GB) tip, and field (LMC disk) bright  main sequence (MS) stars.
The experiments indicated that GB tip and bright field MS stars are the relevant ones
to explain the suspected integrated color shifts. We adopted typical stellar magnitudes
and colors as observed in BV CMDs of the SWB\,VI clusters like NGC\,1978 
(Bomans et al. 1995): 
V = 16.8, (B-V) = 1.70 for a GB tip star, and V = 16.8, (B-V) = -0.10 for an MS star. 
Since CMDs involving the U color are not available for such
LMC clusters, we adopted solar neighborhood (U-B) values from Gunn \& Stryker (1983) and 
reddened them according to the LMC foreground reddening E(B-V) = 0.06 (Mould \& Aaronson, 
1980), resulting in (U-B) = 2.00 for a GB tip star (M1/M2\,III), and (U-B) = -0.45 (B5\,V) for
a field MS star.

The results are shown in Fig. 9. Clusters like SL\,842 and SL\,8 are outlying and faint, 
with V$\approx$ 14.0. Their fields are not significantly
disturbed by bright MS stars,  but their integrated colors can
be shifted from their observed positions as type VII clusters to the type VI domain
by simply adding 4-5 upper GB stars to their relatively underpopulated 
CMDs. H\,7, a brighter  cluster (V = 12.2), lies in a more central field, where
the contribution of LMC disk stars is significant.  The arrows in Fig. 9 show 
the effects of subtracting 3 GB  and adding 3 MS stars. A  combination
of such stars and/or Galactic foreground stars of suitable magnitude and colors
might explain the observed cluster locus. Finally, we show the effect of subtracting two field
MS stars from the massive cluster NGC\,1978. The shift is only 
moderate despite    the large     color differences between the cluster and 
these blue stars. The effects of adding 4 upper GB stars is only to redden both 
colors by 0.01 mag.

Thus, it appears reasonable from these experiments that faint, apparently old 
clusters lying
in the lower right region of the type VII domain could actually be SWB type VI 
IACs with a 
relatively small number of giants, or similarly that an apparent VII in the upper left 
domain is actually a type IV IAC with a relatively large number of bright giants.

Fluctuations also appear to occur in the opposite sense, as discussed by 
Bica et al. (1996): 
the giant branch of the old, faint cluster    
Reticulum is  so sparsely populated that it is located in the SWB\,V region, despite
having an age indistinguishable from that of M3 (Walker 1992a).
Since Reticulum is a metal poor globular cluster (Gratton \& 
Ortolani 1987a, Walker 1992a), the
bright giant used for the experiment must be accordingly metal poor. We used
V=17.5, (B-V)=1.02, (U-B)=0.57 (Gratton \& Ortolani 1987a; Eggen 1972). Thus in
Figure 9 the R arrow associated with Reticulum corresponds to a different star from
that used for the IACs. Reticulum is far out in the LMC where the LMC disk 
main sequence is 
absent. Some bright yellow stars (probably Galactic field) are however
present in that
CMD. We also include in our experiments (Y in Figure 9) this type of star, 
adopting V=16.7,
(B-V)=0.65 and (U-B)=0.27. The (U-B) value is from a Galactic G5\,V star
from Gunn \& Stryker (1983), reddened to the LMC extinction. The experiments in 
Figure 9 show that, in combination with photometric errors, Reticulum's integrated
colors might move to the SWB\,VII zone.

In the present work we have carried out an approach concerning the
influence of a few bright stars on the integrated light of IACs. Specifically, 
we are not
computing a standard deviation from the mean integrated color but we are
simply adding or subtracting the flux produced by selected star types from 
the cluster observed integrated color. Consequently, in comparison with the 
statistical analysis in Girardi et al. (1995), who studied cluster color 
dispersions,
we are studying the color
effects produced by stars in the tail of the Gaussian distribution, which
lead to larger color fluctuations.  The variable observed number
of bright giants in clusters of similar age and mass shows the importance
of the present approach.  We also took into account bright field stars, 
especially those  differing significantly in color with respect to the cluster
mean color.

In addition to stochastic fluctuations,
photometric errors also likely contribute to the presence of
IACs in the type VII area. We show in Figure 9 
total photometric error bars for bright and 
faint clusters from Bica et al. 1996. Observational
errors will be particularly large for fainter clusters in more crowded regions. 

\section{Discussion}

In the most recent discussion of the chemical evolution of LMC clusters,
Olszewski et al. (1991) used
metallicities from the Ca\,II triplet and ages from the literature obtained from CMDs.
Although their metallicities were on a uniform scale, the ages were from a variety
of studies and are not homogeneous.
In the present paper, although no new genuine old clusters were detected,
we have increased the number of IACs with CMDs and accurate ages
by $\approx$50\%,
and also derived
homogeneous relative ages for all LMC clusters with reliable CMDs.
We can thus address the question of the chemical evolution of LMC clusters using an
improved dataset.

We show in Figure 10 the resulting chemical evolution of the LMC
for the IACs and old clusters. The spatial distribution of these 
clusters is shown in Figure 11. The three new clusters (OHSC33, SL126 
and OHSC37) from the CaII spectroscopy with metallicities [Fe/H]$\sim-1.0$
(Table 1) have ages in the range 1-3 Gyr. They fall below (i.e. at
lower metallicities than) the bulk of IACs in the cluster age-metallicity
relation, suggesting that their parent gas clouds have not shared
the same enrichment as that for the IACs located in the upper-left
region of Figure 10. Note that the metallicity of the old cluster
ESO121SC03 is also $\sim -1.0$, which confirms that the LMC has not undergone
important star formation processes during the period from 3--9 Gyrs ago,
and consequently also
not experienced significant chemical enrichment.

The IAC sample with $-1.2<$[Fe/H]$<0.0$ is separated into two metallicity groups
in Figure 10. Their spatial distribution in Figure 11
is well-mixed, suggesting no significant metallicity gradient for the
disk, as also found by Olszewski et al. (1991). Nevertheless, it
is necessary to know  the dynamics of the clusters 
in order to investigate whether there exists any similarity between the
present abundance distribution in the disk and the so-called
paleodistributions, namely, the distribution in metallicity of the
gas out of which the clusters were generated. Note that the LMC IACs'
metallicity range is twice as large as that of Galactic open clusters.
The former were mostly born during an $\sim$ 2 Gyr period, while the formation
of the Galactic open cluster system took  at least 6-8 Gyr (see Fig. 3
of Piatti et al. 1995). This fact suggests that the LMC has suffered more
violent star formation processes than those that occurred during the formation
of the Galactic disk. Some IACs have been born in regions whose chemical
enrichment ocurred on a shorter time scale  than
those for other parts of the LMC disk.
On the other hand, the fact that there exist IACs with different metal
contents within the same spatial region (see Figure 11) suggests
that parts of the LMC were not chemically well-mixed, or that the progenitor
gas clouds of differing metallicity that originated in different regions have
subsequently been brought together by large random motions.

\section {Concluding remarks}
\label{conclu}

We have obtained CMDs from Washington
$C,T_1$ photometry for 25 candidate old clusters in the LMC  
and determined ages from the $T_1$ magnitude difference ($\delta$) between
the giant branch clump and the main sequence turnoff for 23 of these.
The calibration of  age vs. ($\delta T_1$)  was obtained using LMC and Galactic clusters
with well-determined parameters.
Our technique is capable of detecting the main sequence of a 10 Gyr old cluster in an
uncrowded field with the 0.9m  CTIO telescope,
as demonstrated by the present photometry of the ``standard" cluster ESO121-SC03.

The candidates turned out to be of intermediate age (1--3 Gyr), although we cannot 
rule out old age for NGC\,1928 and NGC\,1939,  which are compact 
and are located in crowded bar fields. 
The reason  why integrated photometry can confuse intermediate 
age clusters with genuine old 
ones can be understood by stochastic effects produced by bright cluster giants 
and bright field main sequence stars for clusters fainter than V$\approx$13.5, 
and by photometric errors for faint clusters in crowded fields. 

The present study  constitutes a significant increase in the sample of intermediate age 
clusters in the LMC with CMDs. The determination of relative ages  allows one
to obtain accurate information on the LMC history of cluster formation/disruption 
for ages $>1$\,Gyr. 
Our results confirm that the LMC apparently did not 
form clusters between 3--8 Gyrs ago,
and that it only formed a single (surviving)
cluster between 3 Gyrs and the epoch of formation of the classical globular clusters.
We have also found additional evidence that the onset of IAC formation
ocurred at 3 Gyr.

The relatively metal poor ([Fe/H]$\approx -1.0$) candidates 
from the CaII triplet spectroscopy
also turned out to be of intermediate age. They occupy a locus below 
that of the  bulk of the intermediate age clusters  in the LMC age-metallicity relation 
of Olszewski et al. (1991).  These clusters (SL\,126, OHSC\,33, OHSC\,37)  
are outlyers in the LMC disk, suggesting that 
part of the parent gas clouds in these regions  have not
shared the same chemical enrichment as that in the inner regions, 
or that the outer gas clouds were not chemically well-mixed.
This result points to new interesting scenarios 
to be explored
for formation and evolution models of the LMC. We will readdress
the issue of metallicities in our program clusters in a future paper.

In addition to the eight LMC clusters already established as old by means of deep
CMDs and the presence of RR\,Lyraes, NGC\,1939 and NGC\,1928 are still candidates
together with NGC\,1754, NGC\,1898, NGC 1916,
NGC\,2005 and  NGC\,2019. High spatial resolution  photometry, e.g. with Hubble Space 
Telescope, is necessary for
a definitive diagnosis of these clusters
and indeed such studies are underway. Our investigation of the possible
missing faint old clusters has yielded no new members. It is however possible that such 
clusters do exist but have not yet been identified. 
For example, Wielen (1988), using the results of Mateo (1988b), argues that a typical
dissolution time is $\sim 2$ Gyr for a cluster in the outer LMC with an initial mass
of $500 M_\odot $ and a core radius of 1 pc. Thus, among the large number of faint 
unstudied clusters, several clusters older than $\sim 3$ Gyr may exist.
Our technique of obtaining the CMD directly is quite
efficient for searching for such clusters,
requiring only an hour of time on an 0.9m telescope per cluster, and
no followup observations are
required, at least for clusters not severely affected by crowding.

\acknowledgements


J.J.C. and A.E.P. are grateful to the CTIO staff and personnel for hospitality and  support
during the observing run, and particularly to M. Smith for support after the run during the
reduction process. E. Geisler provided moral support and encouragement.
Mauricio Navarette did an expert job of obtaining the 4m observations as part of a service
observing night. We thank A. Sarajedini and G. Da Costa
for a critical reading of an earlier version
of this manuscript.
This work was partially supported by the Brazilian institutions CNPq and FINEP, the Argentine 
institutions CONICET and CONICOR, and the VITAE and Antorcha foundations.
This research is
supported in part by NASA through grant No. GO-06699.01-95A (to DG) from the Space 
Telescope Science
Institute, which is operated by the Association of Universities for
Research in Astronomy, Inc., under NASA contract NAS5-26555.


\newpage
                     
\centerline{\bf Figure Captions}
\begin{figure}
\figcaption{The T1 (R) frame for the field of SL555. This is from a 15 minute
exposure with the CTIO 0.9m telescope. The field is $13.6\arcmin$ on a side. About
33,000 objects were photometered on this frame.}
\end{figure}                                                     

\begin{figure}
\figcaption{
Washington $T_1$ vs. $(C-T_1)$ CMD for
ESO121-SC03 from a total exposure time of 55 minutes with the
CTIO 0.9m.
All objects within 200 pixels of the cluster center are shown.
An equal-area adjacent field contains typically only 14 objects.
The magnitude difference between HB and turnoff is $\delta T_1=3.0$ (Section 3).}
\end{figure}                                                     

\begin{figure}
\figcaption{Comparison of the magnitude difference  ($\delta$) between clump/HB and 
turnoff in different filters: 
circles are R vs. V, triangles are $T_1$ vs. V, X is  R (Pyxis) vs. V (LMC old 
clusters). Typical error bars are shown.
As reference, we show the line of equal values for the axes (solid line). We show also 
a linear fit (dashed line).}
\end{figure}                                                     

\begin{figure}
\figcaption{$\delta$V vs. turnoff V magnitude. Different SWB types are indicated,
when available. Triangles represent SWB VII clusters from Bica et al. (1996)
which we find to be IACs.}
\end{figure} 

\begin{figure}
\figcaption{Age calibration of the $\delta T_1$ index, i.e. the mag.
difference in $T_1$ between the clump/HB and the turnoff, derived from LMC and
Galactic clusters with good main sequence photometry and accurate ages. The
curve is a 3rd order fit.}
\end{figure} 

\begin{figure}
\figcaption{CMD for H7, a cluster in the middle of the age range of our sample.
Note that $\delta T_1=1.1$. All objects within 100 pixels of the cluster center
have been included. 
An equal-area adjacent field typically has only about 1/3 as many stars, concentrated to
the main sequence.}
\end{figure} 

\begin{figure}
\figcaption{ CMD for SL388, a cluster among the oldest IACs in our sample. 
Note that $\delta T_1=1.9$.
All objects within  75 pixels of the cluster center
have been included.
An equal-area adjacent field typically has only about 1/10 as many stars, concentrated to
the main sequence.}
\end{figure} 

\begin{figure}
\figcaption{Age histograms for the IACs (upper panel) and old clusters 
(lower panel) from Tables 1 and 2.
Filled bins in the upper panel refer to clusters with previous CMDs in 
the literature, while the results from our photometry are shown as the open bins.}
\end{figure} 

\begin{figure}
\figcaption{ Enlargement of Figure 2a from Bica et al. (1996), showing 
the integrated (U-B) vs. (B-V) diagram for LMC
clusters (dots), except: (i) fiducial old clusters (filled squares),
(ii) candidate old clusters not studied in the present sample (asterisks),
(iii) candidates from the present sample (open squares).
 Arrows leading from some of these clusters
show the effects of possible fluctuations produced by small changes in the 
number of bright stars. R represents the addition (or subtraction) of upper giant
branch stars; B field main sequence stars; Y yellow bright stars
(probably Galactic field stars). The error bars show total
photometric errors for our program clusters. Both of these effects together
can account for the presence of IACs in the type VII domain.}
\end{figure} 

\begin{figure}
\figcaption{Chemical evolution of the LMC for IACs (triangles) and old clusters
(circles). The more metal poor IACs are indicated by filled triangles, and the
candidates from Olszewski et al. are labelled.} 
\end{figure} 

\begin{figure}
\figcaption{Spatial distribution of clusters which have both metallicity values
(Olszewski et al. 1991) and ages from CMDs as determined in this paper. As a
reference, the LMC bar is schematically indicated.} 
\end{figure} 


\begin{references}


Barbaro, G., \& Bertelli, G. 1977, A\&A, 54, 243

Battistini, P.L., B\`onoli, F., Casavecchia, M., Ciotti, L., 
Federici, L., \& Fusi Pecci, F. 1993, A\&A, 272, 77

Bica, E., Clari\'a, J.J, \& Dottori, H.
1992, AJ, 103, 1859

Bica, E., Clari\'a, J.J., Bonatto, C., Piatti, A.E., Ortolani, S., 
\& Barbuy, B. 1995, A\&A, 303, 747

Bica, E., Clari\'a, J.J, Dottori, H., Santos, J.F. Jr., \& Piatti, A.E.
1996, ApJS, 102, 57

Bomans, D.J., Vallenari, A., \& de Boer, K.S. 1995, A\&A, 298, 427

Brocato, E., Buonanno, R., Castellani, V., \& Walker, A.R. 1989, 
ApJS, 71, 25

Brocato, E., Castellani, V., Ferraro, F.R., Piersimoni, A.M., \& Testa, V.
   1996, MNRAS, 282, 614

Butcher, H. 1977, \apj , 216, 327

Canterna, R. 1976, AJ, 81, 228

Carraro, G., \& Patat, F. 1994, A\&A, 289, 397

Chiosi, C., Bertelli, G., \& Bressan, A. 1988, A\&A, 196, 84

Corsi, C.E., Buonnanno, R., Fusi Pecci, F., Ferraro, F.R., Testa, V.,
\&   Greggio, L. 1994, MNRAS, 271, 385


Da Costa, G.S. 1991, in The Magellanic Clouds, ed. R. Haynes \&
D. Milne (Dordrecht:Kluwer), 145

Da Costa, G.S. 1995, PASP, 107, 937

Da Costa, G.S., King, C.R., \& Mould, J.R. 1987, ApJ, 321, 735

Da Costa, G.S., Mould, J.R. \& Crawford, M.D. 1985, ApJ, 297, 582

Dinescu, D.I., Demarque, P., Guenther, D.B., \& Pinsonneault, M.H. 
1995, AJ, 109, 2090

Djorgovski, S., Thompson, D.J., de Carvalho, R.R., \& Mould, J.R.
1990, AJ, 100, 599

Dottori, H., Melnick, J., \& Bica, E. 1987, RMxA\&A, 14, 183

Eggen, O.J. 1972, ApJ 172, 639

Elson, R.A.W., \& Fall, S.M. 1985, ApJ, 299, 211

Elson, R.A.W., Gilmore, G.F., \& Santiago, B.X. 1997, \mnras , 289, 157

Flower, P.J., Geisler, D., Hodge, P.W., \& Olszewski, E.W. 1980, ApJ, 235,
   769

Flower, P., Geisler, D., Hodge, P.W., Olszweski, E., \& Schommer, R. 1983, 
   ApJ, 265, 15

Gallagher, J.S., et al. 1996, \apj , 466, 732

Geisler, D. 1987, AJ, 93, 1081

Geisler, D. 1996, AJ, 111, 480

Geisler, D., \& Sarajedini, A. 1996, in Formation of the Galactic 
Halo....Inside and Out, ASP Conf. Ser., no. 92, ed. H. Morrison \& 
A. Sarajedini (San Francisco:ASP), 524

Geisler, D., \& Sarajedini, A. 1997, in preparation

Girardi, L., \& Bica, E. 1993, A\&A 274, 279

Girardi, L., Chiosi, C., Bertelli, \& G., Bressan, A. 1995, A\&A,
298, 87

Gratton, R.G., \& Ortolani, S. 1987a, A\&AS, 71, 131

Gratton, R.G., \& Ortolani, S. 1987b, A\&AS, 67, 373

Gunn, J.E., \& Stryker, L.L. 1983, ApJS, 52, 121

Hardy, E., Melnick, J., \& Reheault, C. 1980, in  Star Clusters, ed. J.E.
   Hesser, (Dordrecht:Reidel), 343

Harris, H.C., \& Canterna, R. 1979, AJ, 84, 1750

Harris, W.E. 1991, ARA\&A, 29, 543

Harris, W.E., Hesser, J.E., \& Atwood, B. 1983, PASP, 95, 967

Harris, W.E. Phelps, R.L., Madore, B.F., Pevunova, O., Skiff, B.A., 
Crute, C., Wilson, B., \& Archinal, B.A. 1997, AJ, 113, 688

Hesser, J.E., McClure, R.D., \& Harris, W.E. 1984, in  Structure and
Evolution of the Magellanic Clouds, ed. S. van den Bergh \& K.S. de Boer, 
(Dordrecht:Reidel), 47

Hodge, P. 1960, ApJ, 131, 351

Hodge, P.W. 1984, PASP, 96, 947

Hodge, P.W., \& Flower, P. 1987, PASP, 99, 734


Irwin, N.J., Demers, S., \& Kunkel, W.E. 1995, ApJL, 453, 21

Janes, K.A., \& Phelps, R.L. 1994, AJ, 108, 1773

Jensen, J., Mould, J., \& Reid, N. 1988, ApJS, 67, 77

Jones, J.H. 1987, AJ, 94, 345

King, I.R. 1966, \aj , 71, 276

Kubiak, M., Kaluzny, J., Krzeminski, W., \& Mateo, M. 1992, AcA, 42, 155

Mateo, M. 1988a, ApJ, 331, 261

Mateo, M. 1988b, in Globular Clusters Systems in Galaxies, ed. J.E. Grindlay
\& A.G. Davis Philip, (Dordrecht:Reidel), 557

Mateo, M., \& Hodge, P.W. 1985, PASP, 97, 753

Mateo, M., \& Hodge, H. 1986, ApJS, 60, 893

Mateo, M., \& Hodge, P.W. 1987, ApJ, 320, 626

Mateo, M., Hodge, P., \& Schommer, R.A. 1986, ApJ, 311, 113

Mighell, K.J., Rich, R.M., Shara, M., \& Fall, S.M. 1996, AJ, 111, 2314

Montgomery, K.A., Janes, K.A., \& Phelps, R.L. 1994, AJ, 108, 585

Montgomery, K.A., Marschall, L.A., \& Janes, K.A. 1993, AJ, 106, 181

Mould, J.R., \& Aaronson, M. 1980, ApJ, 240, 464

Mould, J.R., Da Costa, G.S., \& Crawford, M.D. 1986, ApJ, 304, 265

Mould, J.R., Da Costa, G.S., \& Wieland, F.P. 1986, ApJ, 309, 39

Mould, J., Kristian, J., Nemec, J., Aaronson, M., \& Jensen, J. 1989, ApJ, 
   339, 84

Mould, J.R., Xystus, D.A., \& Da Costa, G.S. 1993, ApJ, 408, 108

Mould, J.R., Xystus, D.A., Da Costa, G.S., \& Schommer, R.A. 1993, ApJ, 416,
582

Olszewski, E.W. 1984, ApJ, 284, 108

Olszewski, E.W. 1988, in Globular Cluster Systems in Galaxies, ed. J.E.
Grindlay \& A.G. Davis Philip (Dordrecht:Kluwer), 159

Olszewski, E.W., Schommer, R.A., Suntzeff, N.B., \& Harris, H.C. 1991, 
AJ, 101, 515

Ortolani, S., Bica, E., \& Barbuy, B. 1993, A\&A, 273, 415

Phelps, R.L., Janes, K.A., \& Montgomery, K.A. 1994, AJ, 107, 1079

Piatti, A.E., Claria, J.J., Abadi, M.G., 1995, AJ, 110, 2813

Santos, J.F.C. Jr., \& Frogel, J. 1997, ApJ, 479, 764

Sarajedini, A., \& Geisler, D. 1996 AJ, 112, 2013

Schommer, R.A., Olszewski, E.W., \& Aaronson, M. 1984, ApJ, 285, L53

Searle, L., Wilkinson, A., \& Bagnuolo, W. 1980, ApJ, 239, 803

Stetson, P.B. 1987, PASP, 99, 191

Stryker, L.L. 1983, ApJ, 266, 82

Suntzeff, N.B., Schommer, R.A., Olszewski, E.W., \& Walker, A.W. 1992, 
AJ, 104, 1743

Testa, V., Ferraro, F.R., Brocato, E., \& Castellani, V. 1995, MNRAS, 275, 454

Tripicco, M.J., Bell, R.A., Dorman, B., \& Hufnagel, B. 1995, AJ, 109, 1697

Vallenari, A., Aparicio, A., Fagotto, F., \& Chiosi, C. 1994, A\&A, 284, 424

Walker, M.F. 1979, MNRAS, 188, 735

Walker, A.R. 1990, AJ, 100, 1532

Walker, A.R. 1992a, AJ, 103, 1166

Walker, A.R. 1992b, AJ, 104, 1395

Walker, A.R. 1993, AJ, 106, 999

Webbink,R.F., 1985, in Dynamics of Star Clusters, ed. J. Goodman \& P. Hut
(Dordrecht:Reidel), 541

Wee, S-O., \& Lee, M.G. 1996, JKASS, 29, 1

Wee, S-O., Lee, M.G., \& Geisler, D. 1996, JKASS, 29, S147

Westerlund, B.E., Linde, P., \& Lyng\aa, G. 1995, A\&A, 298, 399

Wielen, R. 1988, in Globular Clusters Systems in Galaxies, ed. J.E. Grindlay
\& A.G. Davis Philip, (Dordrecht:Reidel), 393

\end{references}
\end{document}


\pagestyle{empty}
\begin{deluxetable}{lcccccccccc}
\tablenum{2}
\tablecolumns{11}
\tablewidth{0pt}
\tablecaption{CMDs From The Literature}
\tablehead{
\colhead{Cluster Name} & \colhead{$V$} & \colhead{SWB} & \colhead{[Fe/H]} & 
\colhead{$V_{TO}$} & \colhead{$R_{TO}$} & \colhead{$\delta{V}$} &
\colhead{$\delta{R}$} & \colhead{CMD} & \colhead{Age(Gyr)} &
\colhead{Age(Gyr)}  \\
\colhead{} & \colhead{} & \colhead{} & \colhead{} & \colhead{} & 
\colhead{} & \colhead{} & \colhead{} & \colhead{ref.} & \colhead{lit.} & 
\colhead{this paper}  } 
\startdata
NGC1866=SL319=LW163                          &    9.73    &    III &  -     & 17.20 &   -   & 1.1 &  -   &  1 & 0.1  & -  \\
NGC1953=SL459                                &   11.74    &    III &  -     &  -    & 17.42 & -    & 0.9 &  3 & 0.25 & -  \\
NGC2010=SL531                                &   11.72    &    III &  -     & 17.10 &   -   & 0.9 &  -   &  4 & 0.12 & -  \\
NGC2031=SL577                                &   10.83    &    III &  -     &  -    & 16.92 & -    & 1.4 & 40 & 0.14 & -  \\
NGC2134=SL760                                &   11.05    &    III &  -     & 17.72 &   -   & 0.5 &  -   &  2 & 0.19 & -  \\
NGC1756=SL94                                 &   12.24    &    IVA &  -     & 17.80 &   -   & 0.3 &  -   &  5 &  -   & -  \\
NGC1831=SL227=LW133                          &   11.18    &    IVA & +0.01  & 18.60 &   -   & 0.1 &  -   &  5 &  -   & -  \\
                                             &            &        &        & 18.92 &   -   & 0.5 &  -   &  6 & 0.40 & -  \\
NGC1868=SL330=LW169                          &   11.57    &    IVA & -0.50  & 19.40 &   -   & 0.3 &  -   &  5 &  -   & -  \\
                                             &            &        &        & 19.33 &   -   & 0.3 &  -   &  7 & 0.70 & -  \\  
NGC2107=SL679                                &   11.51    &    IVA &  -     & 18.25 &   -   & 0.7 &  -   &  5 &  -   & -  \\
NGC1777=SL121=LW96                           &   12.80    &    IVB & -0.35  & 20.10 & 19.70 & 0.6 & 0.7 & 10 & 0.90 & 1.2,1.2\\
NGC2209=SL849=LW408                          &   13.15    &    IVB &  -     & 19.80 &   -   & 0.3 &  -   &  5 &  -   & 1.0\\
                                             &            &        &        & 19.58 &   -   & 0.3 &  -   &  8 & 1.2  & 0.9\\
NGC2249=SL893=LW479                          &   11.94    &    IVB &  -     & 19.14 &   -   & 0.4 &  -   &  2 & 0.55 & 1.0\\
                                             &            &        &        & 19.66 &   -   & 0.5 &  -   &  9 & 0.6  & 1.2\\
NGC1651=SL7=LW12                             &   12.28    &     V  & -0.37: &  -    & 20.25 & -    & 1.6 & 19 & 2.0  & 2.0\\
SL28=LW47                                    &   13.53    &     V  & -0.37: & 20.50 &   -   & 1.1 &  -   & 14 & 2.0  & 1.5\\
SL61=LW79                                    &   13.99    &     V  & -0.50  & 20.74 &   -   & 1.3 &  -   & 18 & 1.8  & 1.8\\
NGC1783=SL148                                &   10.93    &     V  &  -     &  -    & 19.65 & -    & 0.9 & 20 & 0.9  & 1.3\\
NGC1795=SL165                                &   12.42    &     V  & -0.23  &  -    & 19.74 & -    & 1.3 & 12 & 0.9  & 1.7\\
H14=SL506=LW220                              &   13.42    &     V  & -0.66  &  -    & 20.54 & -    & 1.4 & 12 & 2.5  & 1.8\\
H4=SL556=LW237                               &   13.33    &     V  & -0.15  & 20.90 & 20.70 & 1.9 & 2.1 & 11 & 2.0  & 2.8,3.2\\
NGC2162=SL814=LW351                          &   12.70    &     V  & -0.23  & 19.62 &   -   & 0.7 &  -   & 17 & 0.7  & 1.3\\
                                             &            &        &        & 19.80 &   -   & 0.7 &  -   &  5 &  -   & 1.3\\
NGC2190=SL819=LW357                          &   12.94    &     V  & -0.12  & 19.79 &   -   & 0.4 &  -   & 17 & 0.7  & 1.1\\
                                             &            &        &        & 19.90 &   -   & 0.5 &  -   &  5 &  -   & 1.2\\
NGC2193=SL839=LW387                          &   13.42    &     V  &  -     &  -    & 20.30 &  -   & 1.8 & 16 & 2.2  & 2.3\\
NGC2213=SL857=LW419                          &   12.38    &     V  & -0.01  &  -    & 19.85 &  -   & 1.3 & 15 & 1.3  & 1.6\\
NGC2231=SL884=LW466                          &   13.20    &     V  & -0.67  & 20.19 &   -   & 1.0 &  -   & 13 & 1.3  & 1.5\\
SL455=LW207                                  &   14.12    &     VI & -0.77  & 20.31 &   -   & 1.3 &  -   & 14 & 2.0  & 1.7\\
NGC1978=SL501                                &   10.70    &     VI & -0.42  & 20.93 & 20.53 & 1.4 & 1.7 & 23 & 2.2  & 1.9,2.1\\
                                             &            &        &        & 20.47 &   -   & 1.5 &  -   & 24 & 2.0  & 2.0\\
NGC2121=SL725=LW303                          &   12.37    &     VI & -0.61  & 20.07 &   -   & 1.5 &  -   & 22 & 0.4  & 2.0\\
NGC2173=SL807=LW348                          &   11.88    &     VI & -0.24  & 20.32 & 20.25 & 1.6 & 1.7 &39,5& 1.8  & 2.2,2.1\\
NGC2203=SL836=LW380                          &   11.29    &     VI & -0.52  & 20.52 &   -   & 1.4 &  -   & 21 & 1.0  & 1.8\\
NGC2241=SL888=LW471                          &   13.25    &     VI &  -     & 20.42 &   -   & 1.4 &  -   &  9 & 3.5  & 1.9\\
ESO121SC03                                   &   14.04    &    VII & -0.93  & 21.60 &   -   & 3.0 &  -   & 25 & 9.0  & 9.2\\
NGC1466=SL1=LW1                              &   11.59    &    VII & -2.17  & 22.49 &   -   & 3.3 &  -   & 32 &  -   & 12.7\\
RETICULUM=Sersic40/3                         &   14.25    &VII\tablenotemark{*}&  -     & 22.39 &   -   & 3.6 &  -   & 28 &17.0  & 16.7\\
~~~=ESO118SC31&&&&&&&&&&\\
NGC1786=SL149                                &   10.88    &    VII & -1.87  & 22.35 &   -   & 3.5 &  -   & 27 &15.0  & 15.2\\
NGC1841=ESO4SC15                             &   11.43    &    VII &  -     & 22.50 &   -   & 3.3 &  -   & 27 &15.0  & 12.5\\
                                             &            &        &        & 22.31 &   -   & 3.2 &  -   & 29 &  -   & 11.9\\
NGC2210=SL858=LW423                          &   10.94    &    VII & -1.97  & 22.51 &   -   & 3.5 &  -   & 27 &15.0  & 15.8\\
H11=SL868=LW437                              &   11.93    &    VII & -2.06  & 22.52 &   -   & 3.5 &  -   & 26 &15.0  & 15.2\\
NGC2257=SL895=LW481                          &   12.62    &    VII &  -     & 22.20 &   -   & 3.2 &  -   & 38 &13.0  & 11.9\\
                                             &            &        &        & 22.19 &   -   & 3.2 &  -   & 30 &15.0  & 12.1\\
                                             &            &        &        & 22.29 &   -   & 3.5 &  -   & 31 &14.0  & 15.8\\
\tablebreak
SL354=LW177                                  &   13.65    &VII\tablenotemark{**}& -0.08  & 20.49 &   -   & 1.5 &  -   & 14 & 2.2  & 2.0\\
SL842=LW399                                  &   14.15    &VII\tablenotemark{**}& -0.36  & 20.69 &   -   & 1.5 &  -   & 14 & 3.0  & 2.0\\
LW75=SL59w                                   &    -       &    -   &   -    & 19.52 &   -   & 0.6 &  -   & 33 & 0.8  & 1.2\\
LW76=SL59e                                   &    -       &    -   &   -    & 19.57 &   -   & 0.6 &  -   & 33 & 0.8  & 1.2\\
SL869=LW441                                  &    -       &    -   &   -    & 19.91 &   -   & 0.9 &  -   & 34 & 1.5  & 1.4\\
SL34=LW55                                    &    -       &    -   &   -    & 19.92 &   -   & 1.0 &  -   & 33 & 1.0  & 1.5\\
H88-1=NC1                                    &    -       &    -   &   -    & 20.34 &   -   & 1.0 &  -   & 35 & 1.6  & 1.5\\
SL832=LW376                                  &    -       &    -   &   -    & 20.14 &   -   & 1.0 &  -   & 36 & 0.8  & 1.5\\
SL4=LW4                                      &    -       &    -   &   -    & 20.64 &   -   & 1.2 &  -   & 30 & 2.0  & 1.7\\
                                             &    -       &    -   &   -    & 20.16 &   -   & 1.1 &  -   & 35 & 1.4  & 1.5\\ 
LW195=ESO119SC61                             &    -       &    -   & -0.33  & 20.52 &   -   & 1.5 &  -   & 14 & 2.0  & 2.1\\
E2=ESO120SC8                                 &    -       &    -   & -0.70  & 20.76 &   -   & 1.6 &  -   & 37 & 2.0  & 2.2\\
\enddata
\tablenotetext{*}{Candidates from the integrated photometry for old clusters, which CMDs revealed to be intermediate age 
ones.}
\tablenotetext{**}{Genuine old cluster but located in the SWB\,V zone.}
\tablecomments{
 1 Brocato et al. 1989;
 2 Vallenari et al. 1994;
 3 Mould et al. 1993;
 4 Mateo 1988a;
 5 Corsi et al. 1994;
 6 Hodge 1984;
 7 Flower et al. 1980;
 8 Dottori et al. 1987;
 9 Jones 1987;
10 Mateo \& Hodge 1985;
11 Mateo \& Hodge 1986;
12 Jensen et al. 1988;
13 Walker 1979;
14 Olszewski 1988;
15 Da Costa et al. 1985;
16 Da Costa et al. 1987;
17 Schommer et al. 1984;
18 Mateo \& Hodge 1987;
19 Mould et al. 1986;
20 Mould et al. 1989;
21 Harris et al. 1983;
22 Flower et al. 1983;
23 Bomans et al. 1995;
24 Olszewski 1984;
25 Mateo et al. 1986;
26 Mighell et al. 1996;
27 Brocato et al. 1996;
28 Gratton \& Ortolani 1987a;
29 Walker 1990;
30 Hesser et al. 1984;
31 Stryker 1983;
32 Walker 1992b;
33 Westerlund et al. 1995;
34 Walker 1993;
35 Hodge \& Flower 1987;
36 Hardy et al. 1980;
37 Gratton \& Ortolani 1987b;
38 Testa et al. 1995;
39 Mould et al. 1986;
40 Mould et al. 1993}
\end{deluxetable}